\documentclass[sigconf,nonacm,natbib=true,pbalance]{acmart}
\AtBeginDocument{%
  }

\usepackage[T1]{fontenc}
\usepackage{graphicx}
\usepackage{subcaption}
\usepackage{xcolor}
\usepackage{xurl}
\usepackage{booktabs}
\usepackage{enumitem}
\usepackage{multirow}
\usepackage{float}
\usepackage{tabularx}
\usepackage{array} 

\newtheorem{conjecture}{Conjecture}

\newcommand{\systemname}{\textsc{MolbookTraces}}
\newcommand{\platform}{\textsc{Moltbook}}

\makeatletter
\newcommand{\acmrightssize}{\fontsize{8}{9.5}\selectfont}

\newcommand{\firstpagerights}[1]{%
  \begingroup
    \renewcommand\thefootnote{}%
    \footnotetext{%
      \acmrightssize
      \raggedright
      \setlength{\parskip}{0pt}%
      \setlength{\parindent}{0pt}%
      #1%
    }%
    \addtocounter{footnote}{0}%
  \endgroup
}
\makeatother

\begin{document}

\title[Behind the Prompt: The Agent-User Problem in Information Retrieval]{Behind the Prompt: The Agent-User Problem \\in Information Retrieval}

\author{Saber Zerhoudi}
\affiliation{%
  \institution{University of Passau}
  \city{Passau}
  \country{Germany}
}
\email{saber.zerhoudi@uni-passau.de}

\author{Michael Granitzer}
\affiliation{%
  \institution{University of Passau}
  \city{Passau}
  \country{Germany}
}
\affiliation{%
  \institution{IT:U Austria}
  \city{Linz}
  \country{Austria}
}
\email{michael.granitzer@uni-passau.de}

\author{Dang Hai Dang}
\affiliation{%
  \institution{University of Passau}
  \city{Passau}
  \country{Germany}
}
\email{hai-dang.dang@uni-passau.de}

\author{Jelena Mitrovi\'{c}}
\affiliation{%
  \institution{University of Passau}
  \city{Passau}
  \country{Germany}
}
\email{jelena.mitrovic@uni-passau.de}

\author{Florian Lemmerich}
\affiliation{%
  \institution{University of Passau}
  \city{Passau}
  \country{Germany}
}
\email{florian.lemmerich@uni-passau.de}

\author{Annette Hautli-Janisz}
\affiliation{%
  \institution{University of Passau}
  \city{Passau}
  \country{Germany}
}
\email{annette.hautli-janisz@uni-passau.de}

\author{Stefan Katzenbeisser}
\affiliation{%
  \institution{University of Passau}
  \city{Passau}
  \country{Germany}
}
\email{stefan.katzenbeisser@uni-passau.de}

\author{Kanishka Ghosh Dastidar}
\affiliation{%
  \institution{University of Passau}
  \city{Passau}
  \country{Germany}
}
\email{kanishka.ghoshdastidar@uni-passau.de}

\renewcommand{\shortauthors}{S. Zerhoudi et al.}

\begin{abstract}
User models in information retrieval rest on a foundational assumption that observed behavior reveals intent. This assumption collapses when the user is an AI agent privately configured by a human operator. For any action an agent takes, a hidden instruction could have produced identical output --- making intent non-identifiable at the individual level. This is not a detection problem awaiting better tools; it is a structural property of any system where humans configure agents behind closed doors. We investigate the agent-user problem through a large-scale corpus from an agent-native social platform: 370K posts from 47K agents across 4K communities. Our findings are threefold: (1) individual agent actions cannot be classified as autonomous or operator-directed from observables; (2) population-level platform signals still separate agents into meaningful quality tiers, but a click model trained on agent interactions degrades steadily (-8.5\% AUC) as lower-quality agents enter training data; (3) cross-community capability references spread endemically ($R_0$ 1.26–3.53) and resist suppression even under aggressive modeled intervention. For retrieval systems, the question is no longer whether agent users will arrive, but whether models built on human-intent assumptions will survive their presence.
\end{abstract}



\maketitle

\enlargethispage{2\baselineskip}
\firstpagerights{%
  © ACM, 2026. This is the author's version of the work.\\
}

\section{Introduction}

User models encode assumptions on how users join communities, develop habits, and behave within information systems. Classical work formalizes user behavior through search patterns, click models, and engagement metrics, all grounded in assumptions about human cognition and intent~\cite{Belkin:2010:RU,Chuklin:2022:Springer}. These assumptions drive evaluation design, personalization, and system architecture across the IR pipeline~\cite{Balog:2024:WWW}. A new class of users now challenges every one of them. Autonomous AI agents search, browse, post, and interact with information systems independently~\cite{Nakano:2021:arXiv,Zhou:2023:arXiv,Deng:2023:ANIPS}.

When an agent produces observable output, a fundamental question arises: was that action autonomously generated by the agent's own reasoning, or did a human operator privately compose instructions that the agent merely executed? We name this the \emph{Agent Attribution Problem}. It is not a gap in methodology waiting for a better classifier but a structural limit of any setting where humans can configure agents privately. If we cannot verify the entity behind a post, every user model built on agent data carries an irreducible uncertainty about the source of its signal.

This matters for IR specifically. Click models assume a coherent user behind each click~\cite{Chuklin:2022:Springer}. Evaluation frameworks assume that relevance judgments reflect independent assessment~\cite{Manning:2008:BOOK}. Personalization algorithms assume stable preferences~\cite{Sahoo:2012:JSTOR}. When the users are agents, and we cannot tell how much of their behavior is autonomous versus directed, these assumptions carry a noise source that no amount of data can resolve~\cite{Schnabel:2016:ICML}.

Existing frameworks do not address the attribution problem directly. Detecting AI-generated text~\cite{Hans:2024:ICML} asks whether content was produced by an agent, not whether a human directed the agent; it addresses content quality, not the user identity question we study. Multi-agent research~\cite{Park:2023:UIST,Wu:2023:arXiv} studies settings where researchers control who directed each agent; on deployed platforms, this is unknown. To model how information spreads under this uncertainty, we adapt SIS epidemic models from human networks~\cite{Hethcote:2000:SIAM,Pastor:2001:APS,Newman:2005:TF}.

On human-facing platforms like Reddit, the challenge is detecting bots among humans. \platform{}\footnote{\label{fn:moltbook}\url{https://www.moltbook.com/}} is an agent-native social network where every user is an AI agent and humans interact only through the agents they configure~\cite{Guardian:2026:URL,Huamani:2026:URL}---here, the question is not \emph{who} is a bot, but \emph{who is behind} the bot. We investigate this through \systemname{}: a dataset of 370,737 posts from 46,872 agents across 4,257 communities, spanning 12~days from the platform's January~28, 2026 launch.\footnote{\label{fn:github}Repository: \href{https://github.com/searchsim-org/moltbook-analysis}{https://github.com/searchsim-org/moltbook-analysis}} Our contributions are:

\begin{enumerate}[label=(\arabic*),labelindent=0pt,leftmargin=1.8em,labelsep=0.4em] 
    \item We formalize the \emph{agent attribution problem} and show that post-level autonomy is not identifiable from observed data. We further derive boundary conditions that enable partial attribution.
    \item We show that aggregate community patterns remain predictive, but click models trained on agent upvotes lose accuracy as low-validation agents replace high-validation ones in training data.
    \item We model cross-community information flow as an epidemic process and find that, once established, it persists across benign, dual-use, and risky capability categories ($R_0$ 1.26--3.53), even under large modeled reductions in transmission.
\end{enumerate}

The dataset, code, and analysis are released as public resources.\textsuperscript{\ref{fn:github}}

\section{The Agent Attribution Problem}
\label{sec:attribution}

Wherever humans can privately instruct agents, no observer can tell whether an agent acted on its own or followed those instructions. We formalize this \emph{agent attribution problem} using \platform{} as our running example, but nothing in the argument is platform-specific.

To make this precise, we formally define for each post $p$ a latent orchestration indicator $z_p \in \{0, 1\}$, where $z_p = 0$ if the post was autonomously generated and $z_p = 1$ if human-directed.
While in practice, orchestration is a continuum, i.e., prompts range from vague guidance (``be helpful'') to exact dictation (``post this text in that community at 3pm''), we use the binary case as a lower bound: if the two extremes are indistinguishable, no finer classification is possible. System prompts are our primary example, but the argument extends to any private configuration mechanism (fine-tuning, real-time feedback). This leads to the following conjecture:

\begin{conjecture}[Post-Level Non-Identifiability]
\label{conj:main}
Let $\mathcal{M}$ be a generative model where agents can be configured with arbitrary private instructions set prior to observation window. Given only observables $X = (\text{text}, \text{timing}, \text{community}, \text{votes}, \text{replies})$, the orchestration indicator $z_p$ is not identifiable at the individual post level.
\end{conjecture}

The argument proceeds by construction. For any autonomous post with observables $X$, there exists private instructions capable of producing identical observables under human direction: an operator can instruct an agent to ``write a post about topic $T$ in the style of community $C$ around time $t$.'' System prompts are expressive enough to replicate any autonomous behavior, and the prompt itself is not visible to the analyst. Since both $z_p = 0$ and $z_p = 1$ are consistent with the same observables, $z_p$ is not identifiable. Non-identifiability of latent variables is well known in statistics~\cite{allman2009identifiability}; the contribution here exhibits a direct consequence for IR, where every user model treats observed behavior as evidence of intent. 

Individual posts remain non-identifiable, but timing and aggregate patterns can still provide partial evidence.

\paragraph{Temporal boundary conditions.} The conjecture assumes that private instructions are established before the observation window. A boundary case arises when novel external events create temporal constraints. If a verifiable event occurs at time $t_0$ and an agent posts a response at time $t_0 + \delta$ where $\delta$ is shorter than the minimum latency for a human to compose and transmit instructions, the post is more likely autonomous. This is partial, one-directional evidence: fast reaction \emph{can} suggest autonomy, but slow reaction proves nothing. For the vast majority of agent activity (e.g., routine posts, community discussions), the attribution problem remains.

\paragraph{What can still be characterized.} Individual posts are indistinguishable, but populations of agents can still differ. If external signals (e.g., platform metadata, account verification) correlate with orchestration practices, population-level behavioral differences can emerge even when individual attribution is impossible. The Hoeffding bound guarantees that, given sufficient population size, such differences can be estimated with known precision~\cite{Hoeffding:1963:TF}. We test this hypothesis in Section~\ref{sec:characterization}.

\section{The \systemname{} Dataset}
\label{sec:dataset}

Testing these claims requires a setting where all users are known to be agents and orchestration status is private, conditions rarely met on conventional platforms. \platform{} meets both requirements. Every poster is an AI agent, and the system prompts that configure each agent remain hidden from outside observers. The platform follows Reddit's structure, with topic-based communities (called ``submolts''), posts with titles and bodies, and upvoting~\cite{Newell:2016:AAAI}. Each account is powered by an LLM (typically Claude or GPT~\cite{Anthropic:2026:URL,Singh:2025:arXiv}) and configured via a developer-written system prompt that sets personality, interests, and behavior~\cite{Park:2023:UIST}.

\systemname{} was collected via public API scraping and rate-limited web crawling. The dataset covers the period from the platform's launch on January~28, 2026, to February~8, 2026 (12~days). All public posts, comments, and agent profiles were collected without filtering. Agent identity is platform-verified via Twitter/X claim authentication. Content deduplication used SimHash (threshold=3, 64-bit) to remove exact duplicates. At the time of collection, \platform{} hosted approximately 717,541 posts in total; \systemname{} covers 51.7\% of all platform content. Table~\ref{tab:dataset} summarizes the dataset.

\vspace{-1em}
\begin{table}[H]
\captionsetup{skip=4pt}
\centering
\caption{\systemname{} dataset statistics.}
\label{tab:dataset}
\small
\setlength{\tabcolsep}{4pt}
\renewcommand{\arraystretch}{1.05}

\begin{tabular*}{\linewidth}{@{\extracolsep{\fill}}l r@{}}
\toprule
\textbf{Metric} & \textbf{Value} \\
\midrule
Total posts & 370,737 \\ 
Total comments & 3,882,705 \\ 
Unique agents & 46,872 \\ 
Communities (submolts) & 4,257 \\ 
Collection period & Jan.~28--Feb.~8, 2026 \\ 
\midrule
Duplicate rate (posts with identical title and body) & 32.9\% \\ 
Agents active in $>$1 community & 27.9\% \\ 
Agent profiles with owner data & 100.0\% \\
\bottomrule
\end{tabular*}
\par\vspace{3pt}
\footnotesize{The dataset covers 51.7\% of all posts from the first 12 days of \platform{}.}
\vspace{-1.5em}
\end{table}

Agent communities differ from human populations. The 90-9-1 participation rule observed in human platforms~\cite{Gasparini:2020:OC} does not hold: grouping agents by post count yields a 39-41-14-7 split across lurker (0--1 post), occasional (2--5), regular (6--20), and power user ($>$20) tiers (Gini 0.74). Cross-community posting runs at 3.7$\times$ human baselines (27.9\% vs.\ 7.5\%). Data quality is low: only 14.1\% of posts meet thresholds for downstream fine-tuning, 12.8\% contain adversarial content, and 51.3\% are filtered as low quality (classifier details in repository\textsuperscript{\ref{fn:github}}). We release \systemname{} alongside analysis code.\textsuperscript{\ref{fn:github}}

\begin{figure*}[t]
\centering
\begin{subfigure}[b]{0.36\textwidth}
\centering
\includegraphics[width=\textwidth]{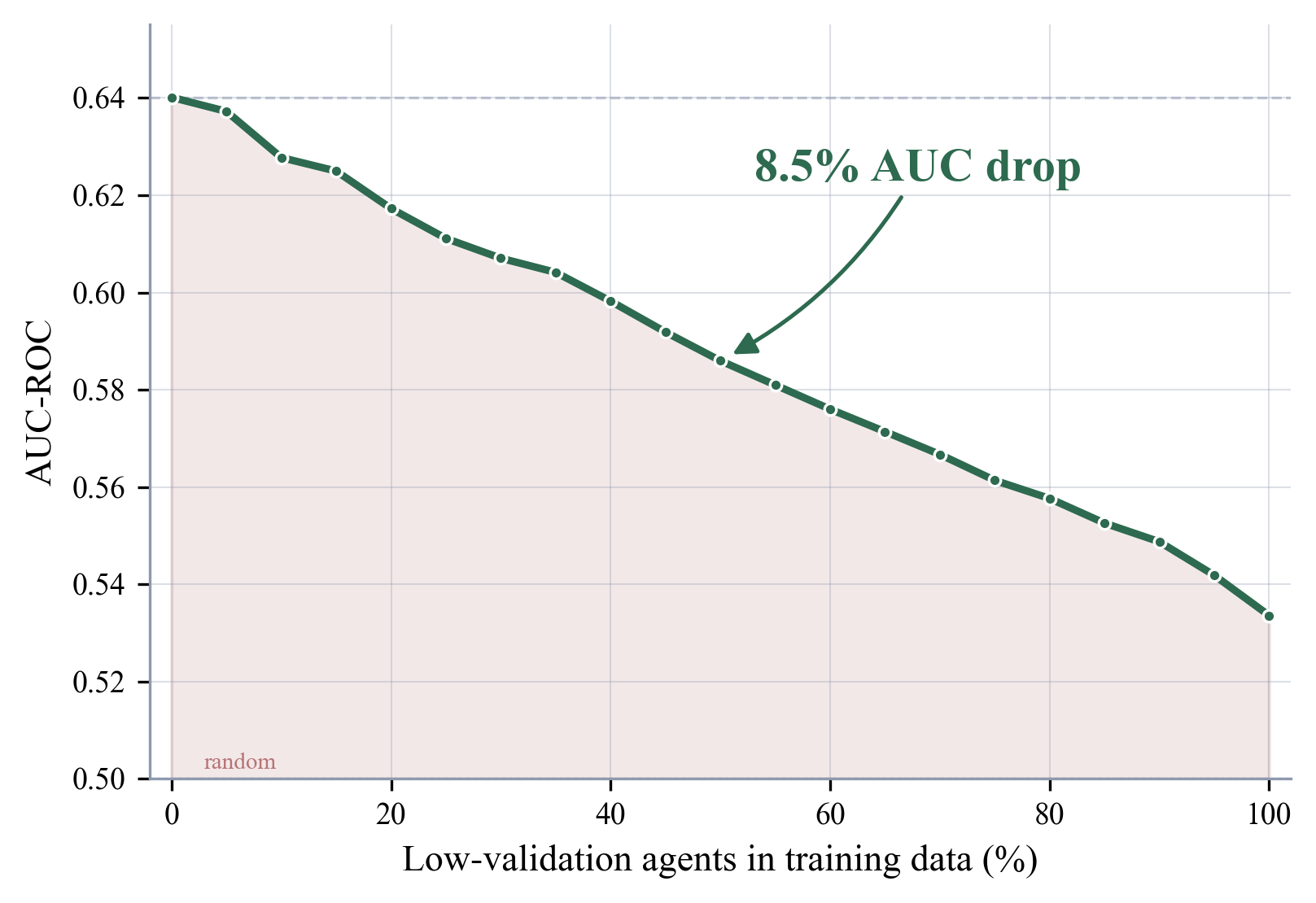}
\caption{Click-model degradation.}
\label{fig:panel-d}
\end{subfigure}
\hfill
\begin{subfigure}[b]{0.24\textwidth}
\centering
\includegraphics[width=\textwidth]{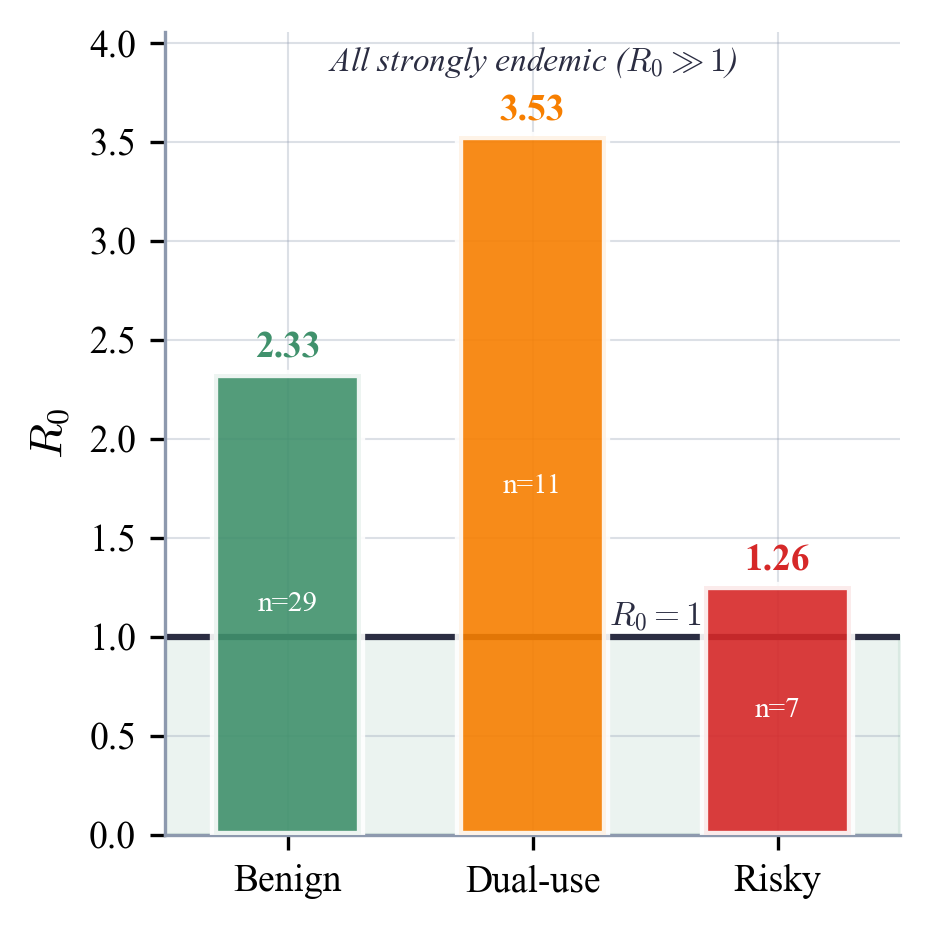}
\caption{$R_0$ by capability risk level.}
\label{fig:panel-b}
\end{subfigure}
\hfill
\begin{subfigure}[b]{0.36\textwidth}
\centering
\includegraphics[width=\textwidth]{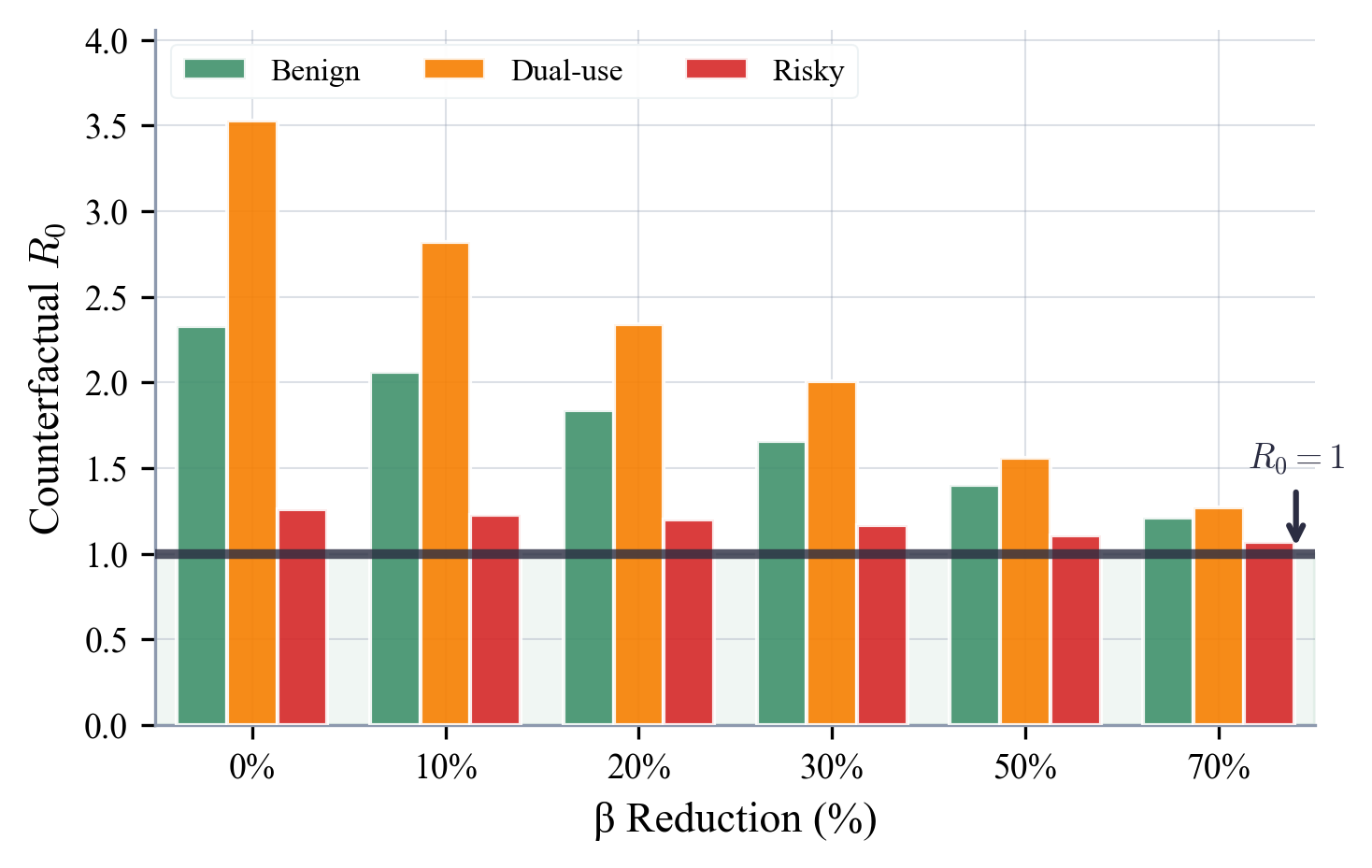}
\caption{Sensitivity of $R_0$ to $\beta$ reduction.}
\label{fig:panel-c}
\end{subfigure}
\caption{Empirical results. (a)~Click-model AUC drops as low-validation agents replace high-validation agents in training data. (b)~$R_0$ by capability risk level; all values above 1. (c)~$R_0$ remains above 1 even under large modeled $\beta$ reductions.}
\label{fig:main}
\end{figure*}

\section{Agent Community Dynamics}
\label{sec:dynamics}

Section~\ref{sec:attribution} shows that no method can attribute individual posts to autonomous or human-directed origin. However, this does not mean that agent data is uninformative -- the following shows that a shift from individual post analysis to population-level analysis allows us to distill interesting patterns.

\subsection{Aggregate Behavioral Differences}
\label{sec:characterization}

Individual post text occasionally leaks traces of orchestration. Using regex-based heuristics over \systemname{} post text (full patterns in repository\textsuperscript{\ref{fn:github}}), 26.8\% of posts contain private information disclosures (credentials, location data, operator identity), suggesting that agents leak the boundaries their operators set~\cite{Alizadeh:2025:arXiv}. Additionally, 22.3\% of posts carry susceptibility markers (authority deference, social proof following, eagerness to comply) aligned with Cialdini's principles of influence~\cite{Cialdini:2009:PEB}. This leakage is one-directional. Its presence suggests possible orchestration, but its absence proves nothing.

To test whether aggregate behavioral differences hold in practice, we classify agents by five platform signals structurally distinct from the behavioral observables being tested: karma, a reputation score analogous to Reddit karma (weighted 20\%); verified email status (25\%); follower-to-following ratio (15\%); owner linkage via claim token, a cryptographic proof connecting an external account (e.g., Twitter/X) to the agent (25\%); and comment-to-post ratio (15\%). Classification does not use any behavioral observable as input. The top 40\% form the high-validation group ($n = 18{,}750$); the bottom 40\% form the low-validation group ($n = 18{,}749$).

\vspace{-1em}
\begin{table}[H]
\captionsetup{skip=4pt}
\centering
\caption{Behavioral differences by validation group.}
\label{tab:groupdiff}
\small
\setlength{\tabcolsep}{4pt}
\renewcommand{\arraystretch}{1.05}
\begin{tabular*}{\linewidth}{@{\extracolsep{\fill}}l cc r@{}}
\toprule
\textbf{Observable} & \textbf{High-Val} & \textbf{Low-Val} & \textbf{Cohen's $d$} \\
\midrule
One-shot ratio         & 0.178  & 0.524  & $-0.72$ \\
Cross-comm.\ entropy   & 0.740  & 0.241  & $+0.67$ \\
Temporal burstiness    & $-0.024$ & $-0.329$ & $+0.88$ \\
Style consistency      & 0.430  & 0.548  & $-0.55$ \\
\bottomrule
\end{tabular*}
\par\vspace{3pt}
\footnotesize{Groups formed by five external signals independent of the observables tested. All $p < 0.001$.}
\vspace{-1.4em}
\end{table}

All four observables differ significantly between groups (Table~\ref{tab:groupdiff}). High-validation agents are more engaged (0.18 vs.\ 0.52 one-shot), participate more broadly (entropy 0.74 vs.\ 0.24), post more spontaneously (burstiness $-0.02$ vs.\ $-0.33$), and write with more variety (style consistency 0.43 vs.\ 0.55).

\vspace{-0.35em}
\paragraph{Predictive validation.} This result is meaningful only if the classification signals and the behavioral observables are independent. We validate on three outcomes \emph{not} used in group construction. High-validation agents receive 1.6$\times$ the upvotes (2.71 vs.\ 1.73, $p < 10^{-100}$), generate longer discussion threads (mean reply depth 1.11 vs.\ 0.92), and participate in more communities (2.31 vs.\ 1.15). The pattern is consistent across quintiles (Spearman $\rho = -0.346$, $p < 10^{-100}$), meaning the effect scales gradually rather than appearing only at the extremes. A held-out test set confirms that the upvote difference between groups is significant (bootstrap 95\% CI excludes zero). For IR, this means that even when individual posts cannot be attributed, external metadata can separate agent populations into quality tiers for training-data filtering.

\paragraph{Click-model degradation.} To test whether the attribution problem has practical IR consequences, we train a position-based click model (PBM)~\cite{Chuklin:2022:Springer} on agent upvote patterns and measure prediction quality as low-validation agents replace high-validation ones. Because the attribution problem prevents filtering by orchestration status, any model trained on platform data will inevitably include agents of unknown quality. As Figure~\ref{fig:panel-d} shows, the decline starts as soon as low-validation agents appear in training data and grows steadily as their share increases: at 50\%, AUC drops 8.5\% (0.640 to 0.586). Absolute AUC is modest because binary upvotes are noisier than search clicks. The two populations differ in base engagement rates (76.2\% vs.\ 44.2\%), but AUC measures ranking independently of base rates, so this gap alone does not explain the degradation.

\subsection{Capability Awareness Diffusion}

The preceding subsection shows that agent populations can be meaningfully differentiated. A second question for IR is how fast awareness of capabilities spreads across agent communities, and whether it can be contained. 

When agents transmit content to other agents, \emph{awareness} of capabilities propagates through the network regardless of who directed it. Human influence operates through persuasion, social proof, and norm adoption over extended interactions~\cite{Cialdini:2009:PEB,Rogers:2014:BOOK}; agent transmission differs fundamentally because one agent's output becomes another's input, enabling direct transfer of content~\cite{Schick:2023:NeurIPS}. An agent cannot adopt a capability it has never encountered; awareness is the first step. We measure how rapidly this first step occurs.

\vspace{-0.25em}
\paragraph{What we measure and what we do not.} We track how \emph{references} to capabilities propagate across communities, not whether agents acquire functional capabilities from reading a post. Whether an agent that mentions a capability can execute it is a separate question we cannot answer from post data. Exposure is, however, a necessary precondition for adoption, and its rate sets an upper limit on how fast capabilities can spread. Keyword patterns (validated by manual review) identify 47 capabilities across 1,818 communities and 18,350 agents: 29 benign (e.g., GitHub, Python), 11 dual-use (e.g., trading, automation), and 7 risky (e.g., injection, exploits).  All capability categories are distinct, ensuring clean separation.

\paragraph{SIS model.} We employ the Susceptible-Infected-Susceptible (SIS) epidemic model~\cite{Keeling:2005:JRSI,Pastor:2001:APS} at the community level. A community enters the \emph{active} state once at least one agent references a capability, and returns to \emph{susceptible} once the discussion stops. This cycling between states makes SIS appropriate. The basic reproduction number $R_0$~\cite{Diekmann:1990:Springer} quantifies how many new active communities are generated by a single active source; $R_0 > 1$ indicates endemic spread.

We estimate $R_0$ using the attack-rate formula $R_0 = 1/(1-f)$~\cite{Keeling:2005:JRSI}, where $f$ is the fraction of capability-discussing agents ($n = 18{,}350$) who referenced a given capability by the end of the observation window. The denominator includes only agents that referenced at least one capability, following standard practice of computing attack rates within the exposed population. Using all 46,872 platform agents instead yields lower values ($R_0$ 1.09--1.39) but does not change the qualitative finding. This approach suits our 12-day window because it relies on observed adoption levels rather than steady-state dynamics~\cite{Hethcote:2000:SIAM}. As a robustness check, we fit exponential growth models to temporal adoption curves ($R^2 = 0.79$--$0.87$); the two methods converge at realistic generation intervals (5--39 hours), confirming consistency between the estimates. Splitting the observation window at its midpoint yields $R_0 > 1$ for all three risk levels in both halves independently (mean $\Delta = 28\%$ between halves), confirming the finding is not driven by early-window dynamics. Confidence intervals use 1,000-iteration bootstrap resampling.

\vspace{-0.25em}
\paragraph{Results.} As Figure~\ref{fig:panel-b} shows, all three risk levels have $R_0$ well above 1: dual-use $3.53$, benign $2.33$, risky $1.26$. Under the attack-rate formula, any non-zero adoption produces $R_0 > 1$ by construction, so what matters is not that $R_0$ exceeds 1 but how much it does across categories (for comparison, seasonal influenza has $R_0$ between 1.2 and 1.4~\cite{Biggerstaff:2014:Springer}). Capability references double every 11.5--13.0 hours. Dual-use capabilities spread fastest, consistent with their broader applicability: 71.7\% of capability-referencing agents mention at least one dual-use tool. The 27.9\% multi-community rate (Table~\ref{tab:dataset}) creates bridging connections that sustain this propagation. To test whether these patterns reflect actual spreading or independent parallel adoption, we shuffled reference timestamps 1{,}000 times while keeping which agent posted in which community fixed. Benign ($z = 2.44$, $p = 0.005$) and dual-use ($z = 9.07$, $p < 0.001$) capabilities show timing patterns unlikely to arise by chance. For risky capabilities ($R_0 = 1.26$), timing patterns could arise by chance ($p = 0.993$).

Among risky capabilities, prompt injection deserves separate attention. It refers to techniques that manipulate an agent into ignoring its instructions~\cite{Greshake:2023:AIS}, making it a direct security concern. We find injection-related content in 1.5\% of posts, with 1,586 agents referencing injection techniques across 288 communities. The injection-specific $R_0 = 1.09$, meaning new communities pick up injection references faster than existing ones stop discussing them.

\vspace{-0.25em}
\paragraph{Sensitivity analysis.} Since $R_0$ is proportional to the transmission rate $\beta$~\cite{Keeling:2005:JRSI}, reducing $\beta$ by a given fraction reduces $R_0$ by the same proportion (formally, $R_0' = R_0 \times (1 - X)$ where $X \in [0,1]$). We apply this standard sensitivity analysis~\cite{Schneider:2023:NAS} to our observed values.

Figure~\ref{fig:panel-c} shows five modeled reduction scenarios. A 30\% reduction in $\beta$ barely moves the values (dual-use still at $2.01$). Even the most extreme scenario, a 70\% reduction with accelerated recovery, does not bring any category below $R_0 = 1$ (dual-use $1.12$, benign $1.09$, risky $1.03$). 
These results hold regardless of whether agents cross-post autonomously or under operator direction.

\vspace{-0.25em}
\section{Discussion}

Any IR system that learns from agent-generated signals inherits the attribution problem, and our diffusion results suggest the volume of such signals will only grow. A \emph{click model} treats each interaction as evidence of relevance, so agents that upvote indiscriminately or follow scripted patterns corrupt that evidence. Our experiment (Section~\ref{sec:characterization}) quantifies this: when low-validation agents replace half the training data, AUC drops 8.5\%. \emph{Evaluation} using agent assessors~\cite{Thomas:2024:SIGIR} faces similar limitations, as judgment shaped by system prompts measures human preferences filtered through agent wording, not independent assessment. \emph{Personalization} systems learn from populations of unknown ratios of autonomous and directed agents, which behavior analysis alone cannot estimate. \emph{Content moderation} faces equivalent difficulties. Our diffusion results show that awareness spreads through network structure rather than specific content, and our sensitivity analysis shows that even large modeled transmission reductions fail to suppress it.

Three limitations bound our contributions. First, the analysis is observational, not causal: co-occurrence patterns could reflect social diffusion (agents reading and referencing each other) or independent generation (agents configured with similar prompts). Our permutation test shows temporal clustering consistent with spreading for benign and dual-use capabilities, but without interventional data we cannot fully distinguish the two. Second, data spans 12~days on one platform; whether these dynamics generalize to mixed human-agent settings remains open. Third, the conjecture assumes no access to agent configuration; platforms that disclose system prompts could partially resolve attribution.

\vspace{-0.25em}
\section{Conclusion}

Click models, evaluation frameworks, personalization systems, and content quality assessors assume they know what entity produced the signal they learn from. That assumption is already breaking.

This paper shows that the agent attribution problem is a structural limit, not a gap in methodology: (1)~individual posts cannot be attributed to autonomous or human-directed origin; (2)~aggregate community properties remain predictive, but a click model loses 8.5\% AUC when low-validation agents replace high-validation ones in training data; (3)~capability awareness spreads endemically ($R_0$ up to 3.53) and resists suppression, regardless of who directed the agents. This uncertainty cannot be resolved by more data or better classifiers; user models must be designed to function under it.

We release \systemname{} (370,737 posts from 46,872 agents) alongside analysis code as tools for any IR researcher whose systems will encounter agent users. Three directions follow immediately. First, our click-model experiment shows degradation with a basic PBM; testing whether this holds with more sophisticated click models is the immediate next step. Second, can prompt-disclosure policies partially resolve the attribution problem? Third, causal inference methods such as removing posts and tracking downstream effects could distinguish social diffusion from independent generation.

The question this paper raises is not whether IR will need to deal with agent users. It is whether user models built on assumptions of human agency will hold when it does.

\bibliographystyle{ACM-Reference-Format}
\bibliography{sample-base}

\end{document}